\documentclass[aps,prl,twocolumn,preprintnumbers,superscriptaddress,10pt]{revtex4-1}

\usepackage{amssymb}
\usepackage{amsmath}
\usepackage{amsfonts}
\usepackage{graphicx}
\usepackage{subfigure}
\usepackage{color}
\usepackage{dcolumn}
\usepackage{hyperref}
\newcommand{\be}{\begin{equation}}
\newcommand{\ee}{\end{equation}}
\newcommand{\bea}{\begin{eqnarray}}
\newcommand{\eea}{\end{eqnarray}}

\begin{document}

\title{Non-equilibrium dynamics and phase transitions }

\author{Romuald A. Janik}
\email{romuald@th.if.uj.edu.pl}
\author{Jakub Jankowski}
\email{jakubj@th.if.uj.edu.pl}
\author{Hesam Soltanpanahi}
\email{hesam@th.if.uj.edu.pl}
\affiliation{Institute of Physics, Jagiellonian University, Lojasiewicza 11, 30-348  Krak{\'o}w, Poland}

\begin{abstract}
We study the poles of the retarded Green's functions of strongly coupled
field theories exhibiting a variety of phase structures
from a crossover up to a first order phase transition.
These theories are modeled by a dual gravitational description. 
The poles of the holographic Green's functions appear at the
frequencies of the quasinormal
modes of the dual black hole background.
We establish that near the transition, in all cases considered, 
the applicability of a hydrodynamic description
breaks down already at lower momenta than
in the conformal case. We establish the appearance of the spinodal region in the case 
of the first order phase transition at temperatures for which the speed of sound squared is negative.
An estimate of the preferential scale attained by the unstable modes is also given.
We additionally  observe a novel diffusive regime for sound modes for a range of wavelengths.
\end{abstract}

\maketitle

\noindent {\bf Introduction.}
One of the most surprising discoveries in contemporary theoretical
physics, the Anti-de-Sitter/Conformal Field Theory (AdS/CFT) 
correspondence  \cite{Maldacena:1997re}, provides for us a way
to investigate the dynamics of 
strongly coupled quantum field theories by means of General Relativity methods.
An important field of research exploiting this new relation between 
geometry and physics aims at exploring the real time dynamics of
strongly interacting hot matter \cite{CasalderreySolana:2011us}.
In particular, real time response of a thermal equilibrium state has been quantified in the case of $\mathcal{N}=4$
super Yang-Mills theory by means of the poles of the retarded Green's function \cite{Kovtun:2005ev}.
The locations of these poles correspond to quasinormal mode (QNM) frequencies in the dual gravitational 
theory. Initial steps towards the extension of this case to nonconformal field theories which 
still admit a gravitational dual description
were taken in ref. \cite{Janik:2015waa,Buchel:2015saa}
where QNM frequencies of an external scalar field were studied.

In this letter we take another step in quantifying real time response of a strongly coupled \emph{nonconformal}
field theory. Firstly, we analyze all allowed channels of energy-momentum tensor perturbations and corresponding 
two-point correlation functions. Secondly, we concentrate on the phenomena appearing in the vicinity
of a nontrivial phase structure of various type: a crossover (motivated by QCD), a $2^{\rm nd}$ order phase transition, a $1^{\rm st}$ order  
phase transition. These cases are modeled by choosing appropriate scalar field self-interaction potentials
in a holographic gravity-scalar theory used in \cite{Gubser:2008ny}. We focus here on phase transitions 
postponing more details of the other cases to \cite{Janik:2016yt}.

Due to the ubiquity of employing a hydrodynamic description for quark gluon plasma it is important to
understand its limitations and the influence of other nonhydrodynamic degrees of freedom.
Let us emphasize, however, that in the present investigation we are targeting quite different physics
from `early thermalization' that was so far predominantly studied within the AdS/CFT correspondence.
Here we expect that the plasma has already thermalized, which could occur at higher temperatures e.g.
in the almost conformal regime. Indeed, the investigations of \cite{Janik:2015waa,Buchel:2015saa} indicate
that nonconformality should not influence much this physics. Subsequently the plasma cools and approaches
a phase transition of appropriate type or a crossover. It is this final stage of plasma evolution that we are considering in
this letter and in particular the pattern of excitations of such a plasma system and the range of applicability
of a hydrodynamic description in the vicinity of a phase transition.

\noindent {\bf The background.}
The black hole background solutions for the QNM
calculations follow from the action
\begin{equation}
S=\frac{1}{2\kappa_5^2}\int d^5x \sqrt{g}  \left[ R-\frac{1}{2}\, \left( \partial \phi \right)^2 - V(\phi) \, \right]~,
\end{equation}
where $\kappa_5=\sqrt{8\pi G}$ and $V(\phi)$ is thus far arbitrary.
These solutions are similar to those studied in ref. \cite{Gubser:2008ny},
but since our goal is to determine the quasinormal mode  frequencies, it will 
be convenient to express them in Eddington-Finkelstein coordinates, which 
have been proven useful in the case of the scalar field modes \cite{Janik:2015waa}. 
The line element reads
\begin{equation}
d s^2 = e^{2 A(r)} (-h(r) d t^2 + d \vec{x}^2) - 2 e^{A(r)+B(r)} d t d r~,
\end{equation}
and for the scalar field $\phi(r)=r$.
Since we require that asymptotically the geometry is that of $AdS$ space
the potential needs to have the following small $\phi$ expansion
\be
V(\phi)= -\frac{12}{L^2} + \frac{1}{2}m^2\phi^2 + O(\phi^4)~. 
\ee
\begin{table}
\centering
\begin{tabular}{c c c c c c c}
\hline \hline
\hspace{5pt}potential \hspace{5pt}& \hspace{5pt} $ a $ \hspace{5pt} & \hspace{5pt} $ \gamma $ \hspace{5pt} & \hspace{5pt} $ b_2 $ \hspace{5pt} & \hspace{5pt} $ b_4 $ \hspace{5pt} & \hspace{5pt} $ b_6 $ \hspace{5pt}  &  $ \Delta $  \\
\hline
\hspace{5pt}$V_{\rm QCD} $ \hspace{5pt}& 0 & \hspace{5pt}0.606 \hspace{5pt} & \hspace{5pt} 1.4 \hspace{5pt} & \hspace{5pt}-0.1 \hspace{5pt} & \hspace{5pt}0.0034\hspace{5pt} &\hspace{2pt}  3.55\hspace{5pt}  \\
$V_{2\rm nd} $ &0  & $1/\sqrt{2} $ &  1.958 & 0  & 0 &  3.38  \\
$V_{1\rm st} $ & 0 & $\sqrt{7/12}$  &  2.5  & 0  & 0 & 3.41  \\
$V_{\rm IHQCD} $ & 1 & $\sqrt{2/3}$ & 6.25 & 0 & 0 & 3.58 \\

\hline
\end{tabular}
\caption{Potentials chosen to study different equations of state
exhibiting different phase structure and corresponding conformal dimension of the scalar field.
\label{t.pot}}
\end{table}

\begin{figure}[t]
\includegraphics[width=.23\textwidth]{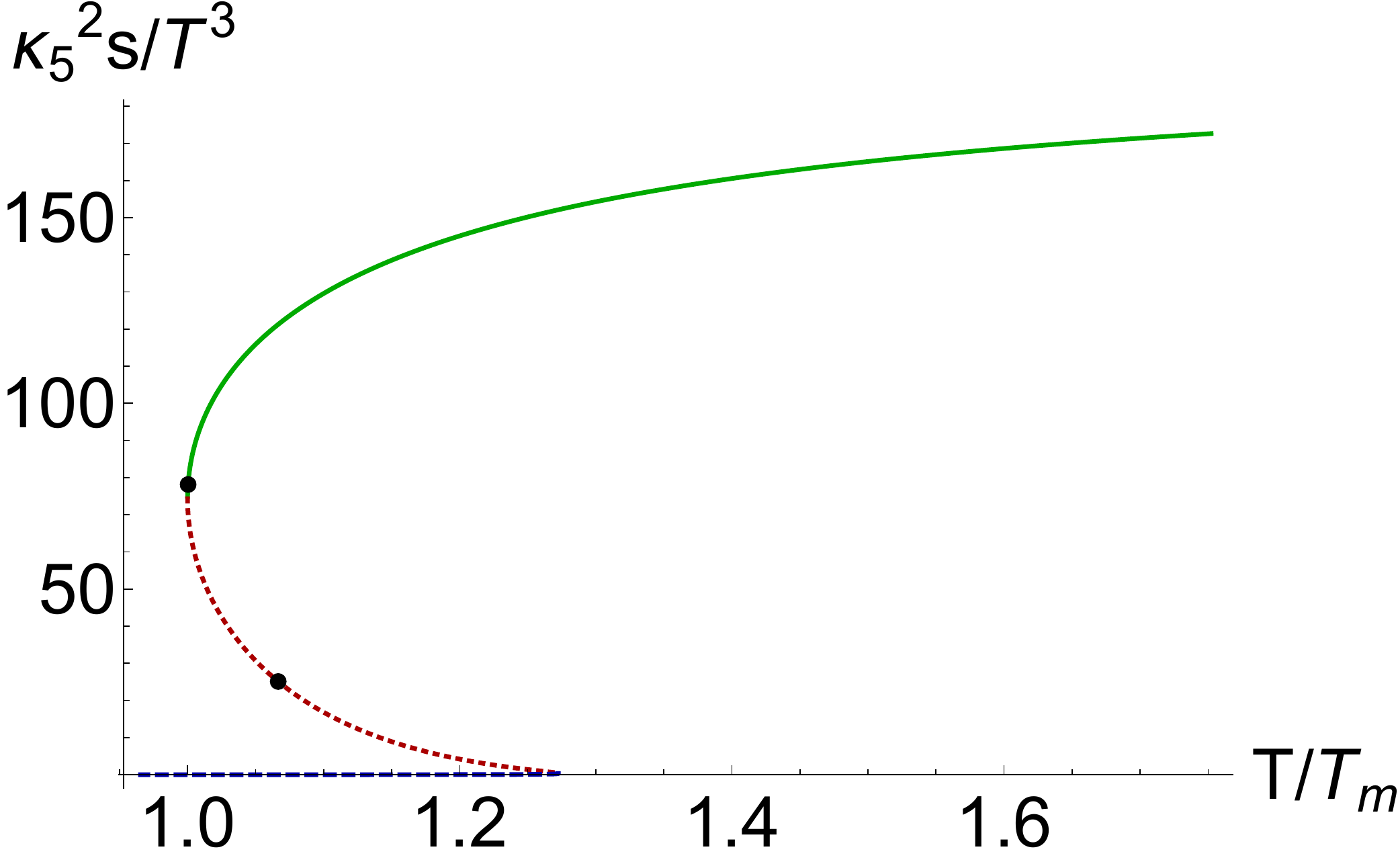}
\includegraphics[width=.23\textwidth]{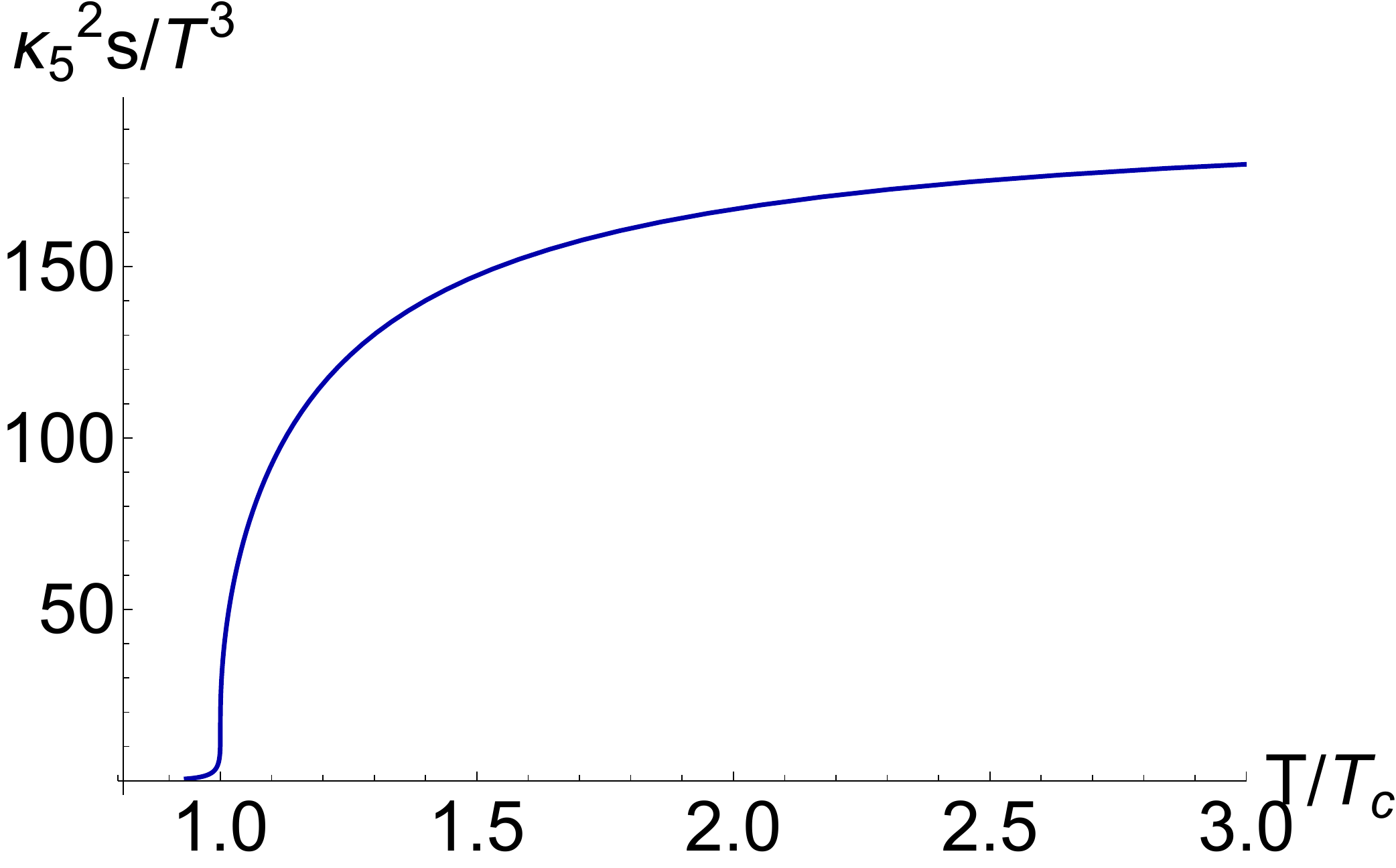}
\caption{Left panel: equation of state for $V_{1\rm st}$ potential with 
different regions marked. Green solid line - stability region.
Red dashed line - spinodal region. QNM's are discussed for solutions
corresponding to black dots. The point where red dashed and green solid lines meet
defines the corresponding minimal temperature
$T_m$ of the unstable branch. Right panel: equation of state for $V_{2\rm nd}$. 
The system exhibits a $2^{\rm nd}$ order phase transition at
$T=T_c$ which is defined by $c_s(T_c)=0$ \cite{Gubser:2008ny}. The specific heat critical
exponent is $\alpha\simeq0.658$.
}
\label{V1st(EoS)}
\end{figure}

Here $L$ is the $AdS$ radius, which we set $L=1$ by the 
freedom of the choice of units. The relation of the scalar mass
and corresponding operator conformal dimension is
$\Delta(\Delta-4)=m^2$.
In general we consider a 
family of potentials \cite{Gubser:2008ny,Gursoy:2008za}:
\begin{equation}
V(\phi)=-12(1+a\phi^2)^{1/4}\cosh(\gamma\phi)+\sum_{n=1}^{3}b_{2n}\phi^{2n}~,
\label{eq:V}
\end{equation}
with parameter values shown in tab. \ref{t.pot}.
The  $a=1$ case is called the Improved Holographic QCD (IHQCD)
\cite{Gursoy:2008za}.
We are interested in solutions possessing a horizon, which requires that the
function $h$ should have a zero at some $\phi=\phi_H$, i.e., $h(\phi_H)=0$.
We solve the coupled equations of motion 
using spectral
discretization and the Newton-Raphson iterative
algorithm. The corresponding equations of state are obtained
from standard procedures in holography. The entropy $s(\phi_H)$ is obtained
from $1/4$ of the area of the horizon, while the temperature $T(\phi_H)$
from the nonsingularity of the euclidean horizon. Consequently one gets
the equation of state $s(T)$, cf. fig.\ref{V1st(EoS)}.




\noindent {\bf  Quasinormal modes.}
We consider perturbations of the background in the following form
$g_{ab}(r,z) = g^{(0)}_{ab}(r) + h_{ab}(r)e^{-i\omega t + i k z}$,
$\phi(r,z) = r + \psi(r)e^{-i\omega t + i k z}$.
Following \cite{Kovtun:2005ev,Janik:2015waa} we consider infinitesimal
diffeomophism transformations, $x^a\mapsto x^a +\xi^a$, 
of the form $\xi_a=\xi_a(r)e^{-i\omega t + i k z}$, 
and look for linear combinations of metric and scalar perturbations which are invariant under
those transformations.  There are four families of such modes, two of which are 
decoupled and two coupled.
Written explicitly, the coupled modes read
\begin{eqnarray}
Z_1(r) & = & H_{aa}(r) \left(\frac{k^2 h'(r)}{2 A'(r)}+k^2 h(r)-\omega ^2\right)\\
\nonumber
&+&k^2 h(r) H_{tt}(r)+\omega  (2 k H_{tz}(r)+\omega  H_{zz}(r))~,\\
Z_2(r) &= & \psi (r)-\frac{H_{aa}(r)}{2 A'(r)}~.
\end{eqnarray}
In the above $h_{aa}(r)=h_{xx}(r)=h_{yy}(r)$ are transverse metric components and 
we have factorized the background from the metric perturbations in the following 
way
$h_{tt}(r) = h(r)e^{2A(r)}H_{tt}(r),$ $h_{tz}(r) = e^{2A(r)}H_{tz}(r),$
$h_{aa}(r) = e^{2A(r)}H_{aa}(r),$  $h_{zz}(r) = e^{2A(r)}H_{zz}(r).$
Comparing with ref. \cite{Kovtun:2005ev} we can see
that $Z_1(r)$ mode corresponds to the sound mode, while the $Z_2(r)$ 
might be called a non-conformal mode, since it is intimately related to the 
scalar field.
The third decoupled mode is the shear one and is expressed as 
$
Z_3(r) = H_{xz}(r) + (\omega/k)H_{tx}(r)
$.
The dynamics of the fourth mode is governed by an equation
of motion which is similar to the external massless scalar 
equation, which was studied in \cite{Janik:2015waa}.
As usual at the horizon we take the ingoing boundary conditions,
which in our coordinates means a regular solution.
The conformal boundary ($r\sim0$) asymptotic is
\be
Z_1(r)\sim A_1 + B_1 r^{\frac{4}{4-\Delta}}~,
\hspace{10pt}
Z_2(r)\sim A_2 r + B_2 r^{\frac{\Delta}{4-\Delta}}~.
\ee
Transformation to the usual Fefferman-Graham coordinates close to the boundary, $r\mapsto\rho^{1/(4-\Delta)}$, reveals 
that $Z_1(\rho)$ has the asymptotics of a massless scalar 
field like the perturbations considered in \cite{Kovtun:2005ev}.
This perturbation corresponds to the sound mode of the theory.
On the other hand $Z_2(\rho)$ has the asymptotics of the 
background scalar field $\phi$ and is similar to the 
case studied in \cite{Benincasa:2005iv}.
According to the AdS/CFT dictionary the boundary conditions are the requirement of vanishing sources,
i.e., $A_1=A_2=0$.
The shear mode perturbation $Z_3(r)$ has the 
same asymptotics as $Z_1(r)$ and requires a 
standard Dirichlet boundary condition at $r=0$.

The problem of determining the quasinormal frequencies
is a form of a generalized eigenvalue
equation, which for given $k$ results in a well defined set of
frequencies $\omega(k)$ \cite{Janik:2016yt}. Note, that all modes, for which 
$\textrm{Re }\omega(k)\neq0$, come in
pairs, i.e., $\omega(k) =\pm|\textrm{Re }\omega(k)| + i \textrm{Im }\omega(k)$. 

\noindent {\bf  Results.}
For all the potentials we have made natural consistency checks.
For high temperatures, i.e., small horizon radius in the sound
and the shear channels we have an agreement with the conformal results 
of ref. \cite{Kovtun:2005ev}. An important thing to note here is that
due to the coupled nature of the modes $Z_1(r)$ and $Z_2(r)$
all frequencies , except for the hydrodynamical one, come in pairs.
This effect is present even at high temperatures, where the 
system is expected to be conformal.  The second most damped nonhydrodynamic
mode turns out to be the most damped one found in ref. \cite{Kovtun:2005ev}.


\begin{table}[ht]
	\centering
	\begin{tabular}{c| c | c}
		\hline \hline
	\hspace{10pt}	potential \hspace{10pt} & \hspace{10pt}  sound channel $q_c$ \hspace{10pt}   & \hspace{10pt} shear channel $q_c$ \hspace{5pt}  \\
		\hline
		$V_{\rm QCD} $ & $0.8$  &  $1.1$    \\
		$V_{2\rm nd} $ & $0.55$ &  $0.9$   \\
		$V_{1\rm st} $ & $0.8$  &  $1.15$  \\
        $V_{\rm IHQCD} $ & $0.14$  &  $1.25$  \\		
		\hline
	\end{tabular}
	\caption{The momenta for which the crossing phenomena between the hydrodynamic  and 
		first nonhydro QNM  happens. Values given at corresponding critical temperatures ($T_{\rm m}$ for $V_{1{\rm st}}$ and $V_{\rm IHQCD}$).	
		\label{t.qc}}
\end{table}


The hydrodynamical QNMs are defined by the condition
$\lim_{k\rightarrow0}\omega_H(k)=0$, and are related
to transport coefficients in the following way
$
\omega\simeq-i\frac{\eta }{s T}k^2,
\omega\simeq\pm c_s k - i\Gamma_s k^2
$,
respectively in the shear and sound channels.
Those formulas are approximate in the sense that
in general higher order transport coefficients
should be considered \cite{Heller:2013fn}.
However, for appropriately small momenta, second
order expansion is enough, and we use it to read off
the lowest transport coefficients of the model.
The sound attenuation constant, $\Gamma_s$,
is related to shear $\eta$ and bulk $\zeta$
viscosities by $\Gamma_s = \frac{\zeta+4\eta/3}{2 s T }$.
Also these formulas were used to make a
second check of the results: compute the speed of sound $c_s$ from the hydrodynamic 
mode and values of the shear viscosity and compare it respectively 
to the ones obtained from the gravitational background calculations and predictions
known in the literature \cite{Kovtun:2004de,Janik:2006ft}. Both of them
are always satisfied, and $\eta/s=1/4\pi$ in all cases considered in this
paper.


\begin{figure}[t]
	\includegraphics[height=.2\textheight]{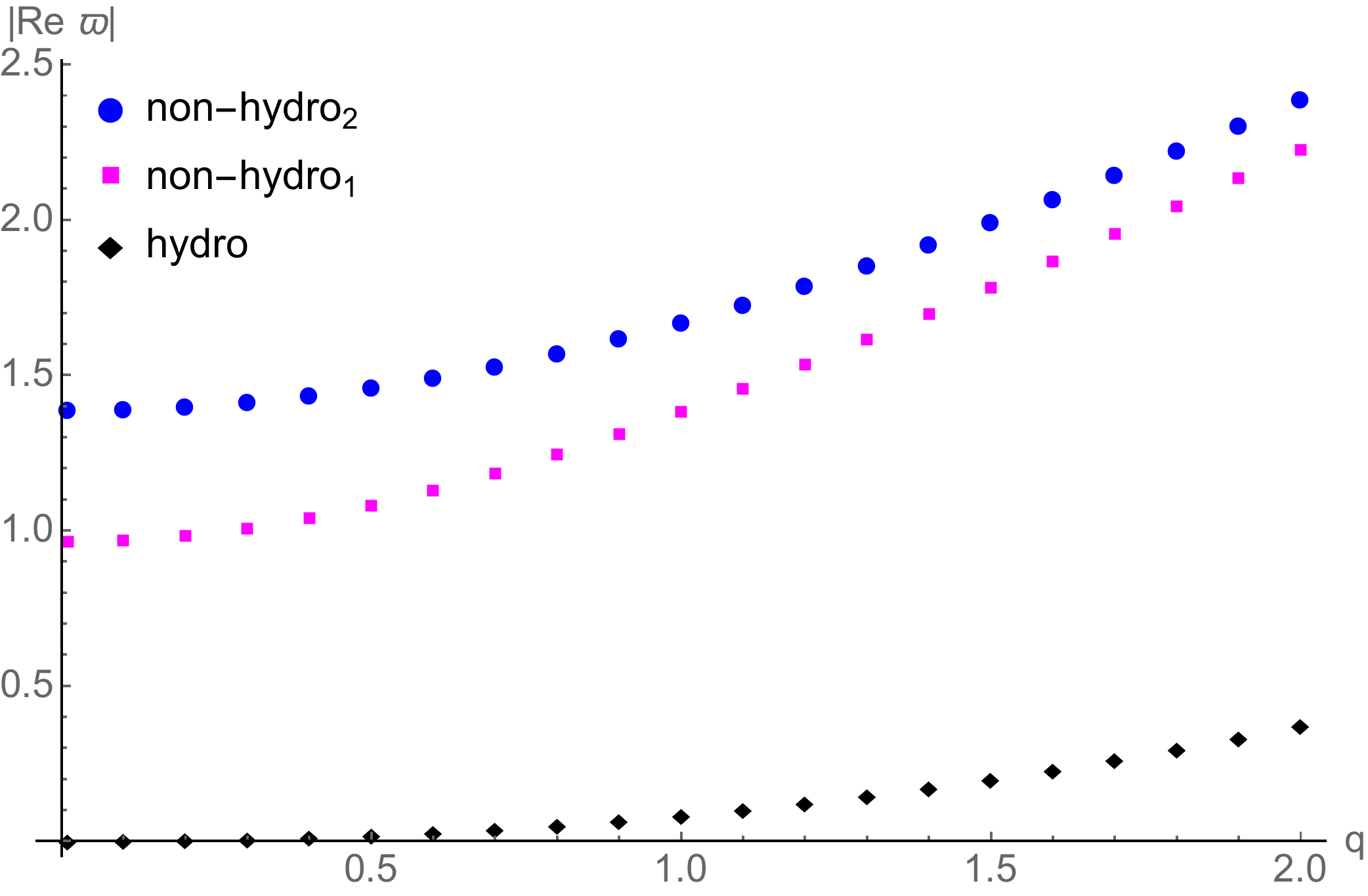}
	\includegraphics[height=.2\textheight]{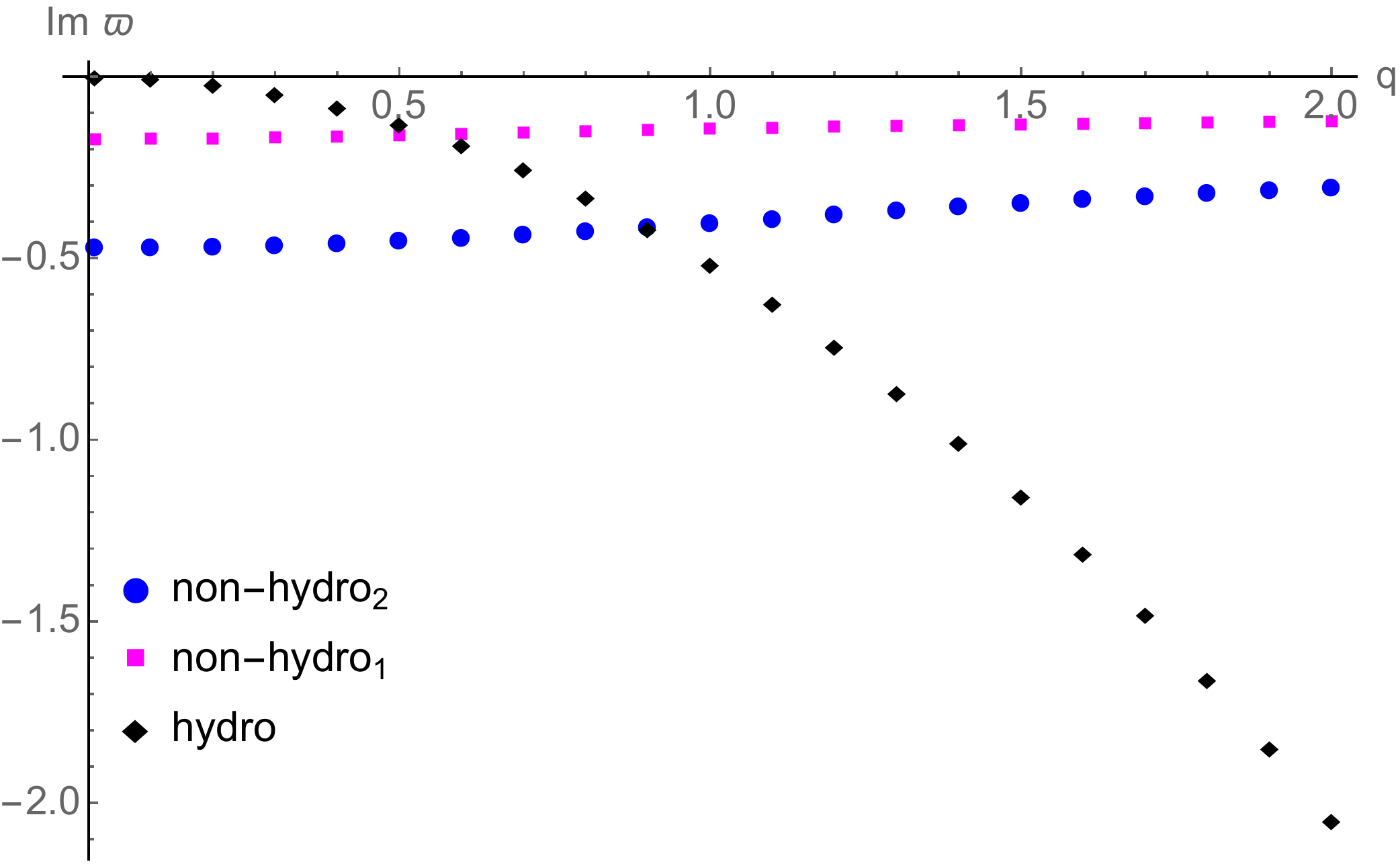}
	\caption{Quasinormal modes for the potential $V_{2\rm nd}$ at $T_c$.
		Real part (upper panel) and imaginary part (lower panel).
		The speed of sound and the ratio of bulk viscosity to the entropy density 
		calculated, namely, $c_s\simeq0,~ \zeta/s\simeq0.061$ (cf.\cite{Gubser:2008yx}).
	}
	\label{V2nd(QNM)}
\end{figure}


In the analysis below we measure the momentum and the frequency in the units of temperature
by setting 
$q=\frac{k}{2\pi T}, 
\varpi= \frac{\omega}{2 \pi T}$.
There are a few novel observations which we make from the pattern of QNM
frequencies. The first is an estimate of the momentum, or equivalently the length
scale, at which the hydrodynamic description of the plasma system breaks down.
For the CFT case this was estimated to be $q=1.3$ where in the shear 
channel first nonhydro QNM dominated the system dynamics \cite{Landsteiner:2012gn}
being less damped than the hydrodynamic shear mode.
The new effect we find
is that away from conformality, in the vicinity
of a phase transition, we see this crossing not only
in the shear channel but also in the sound channel, as illustrated in fig.  \ref{V2nd(QNM)}. 
This shows that 
the influence of a non-trivial phase structure of the background 
affects the applicability of hydrodynamics in a qualitative way. 
Moreover, the momentum at which this crossover happens in the sound channel
is smaller 
which means that the applicability of hydrodynamics 
near the phase transition is more restricted than in the high 
temperature case. In tab. \ref{t.qc} we summarize the critical values of momenta where the hydrodynamic description of the system breaks in sound and shear channels  for the potentials we considered in this letter.


\begin{figure}[b!]
	\includegraphics[height=.2\textheight]{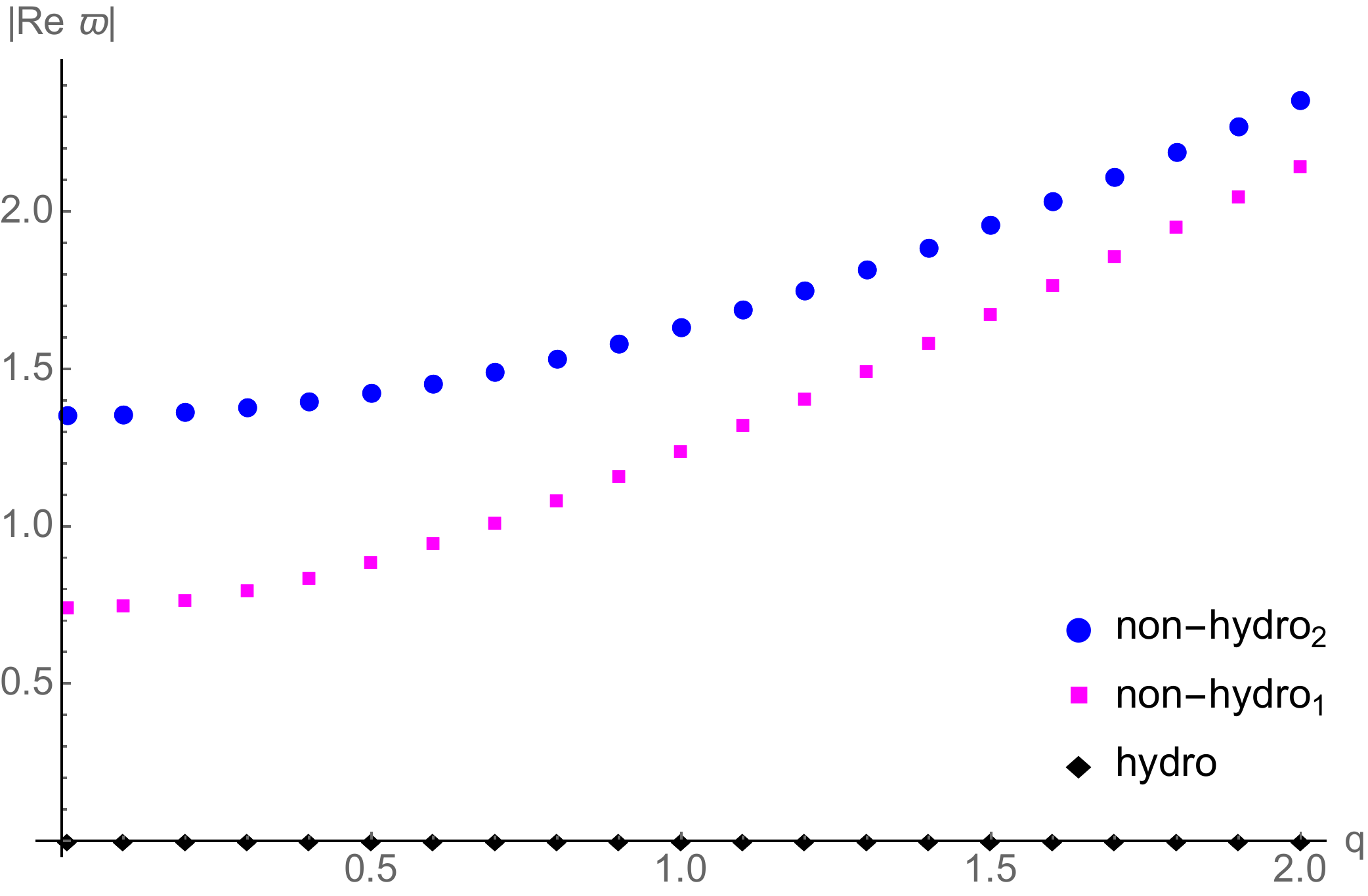}
	\includegraphics[height=.2\textheight]{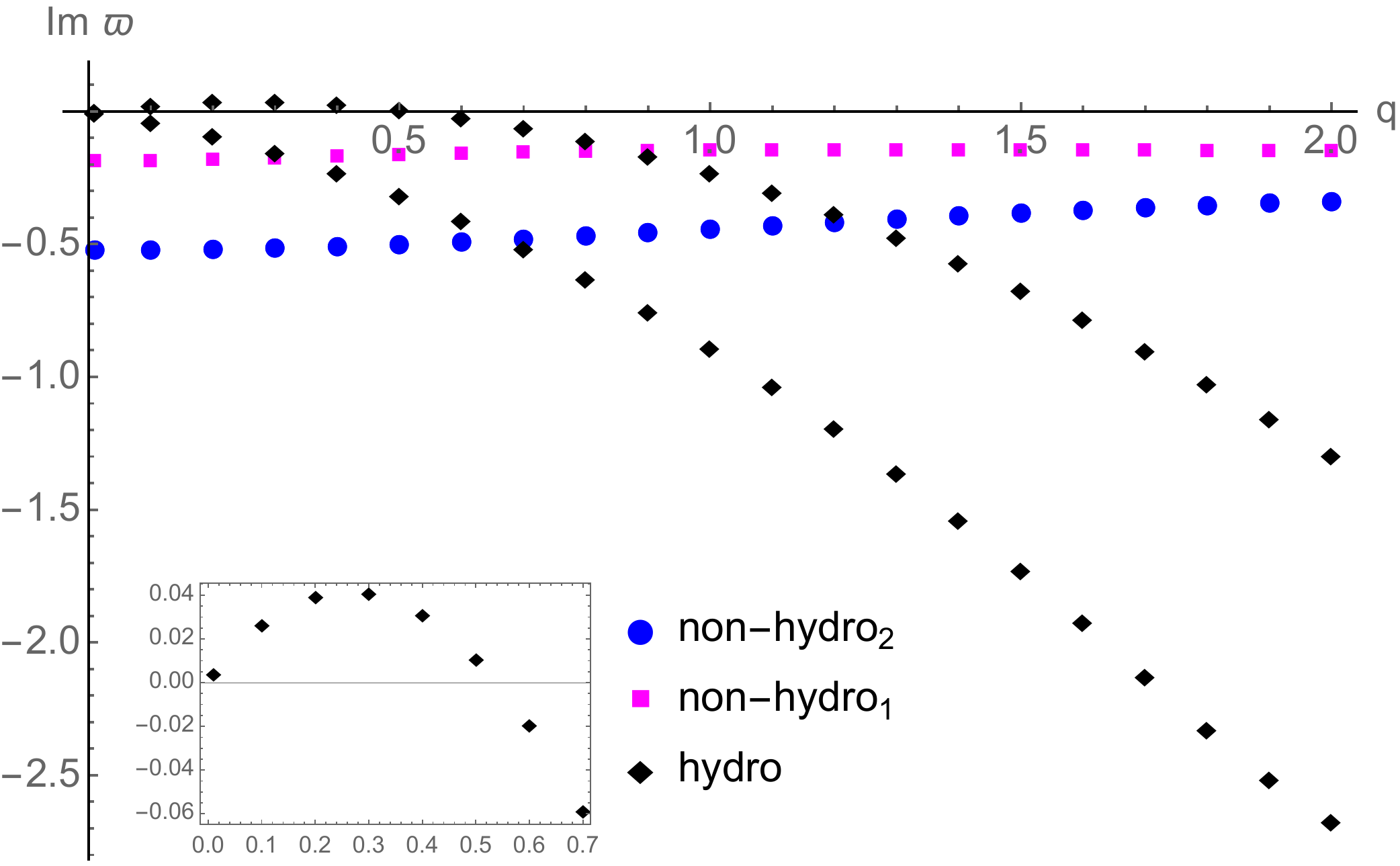}
	\caption{Sound channel quasinormal modes for the potential $V_{1 \rm st}$ at  $T\simeq 1.067 T_m$.
		An instability of the spinodal region is shown in the insertion.
	}
	\label{V1st(QNM)}
\end{figure}

\begin{figure}[h!]
	\includegraphics[height=.2\textheight]{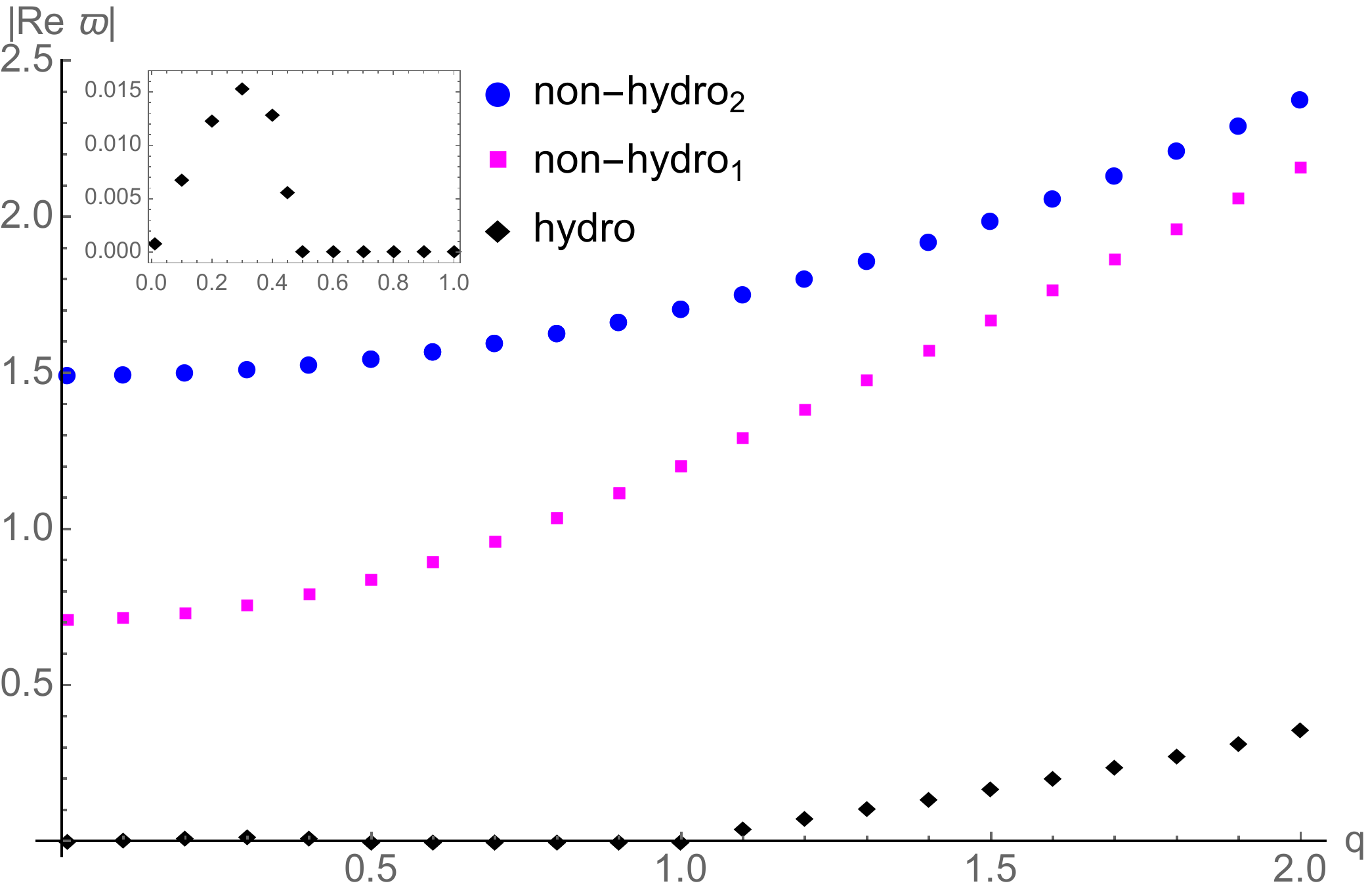}
	\includegraphics[height=.2\textheight]{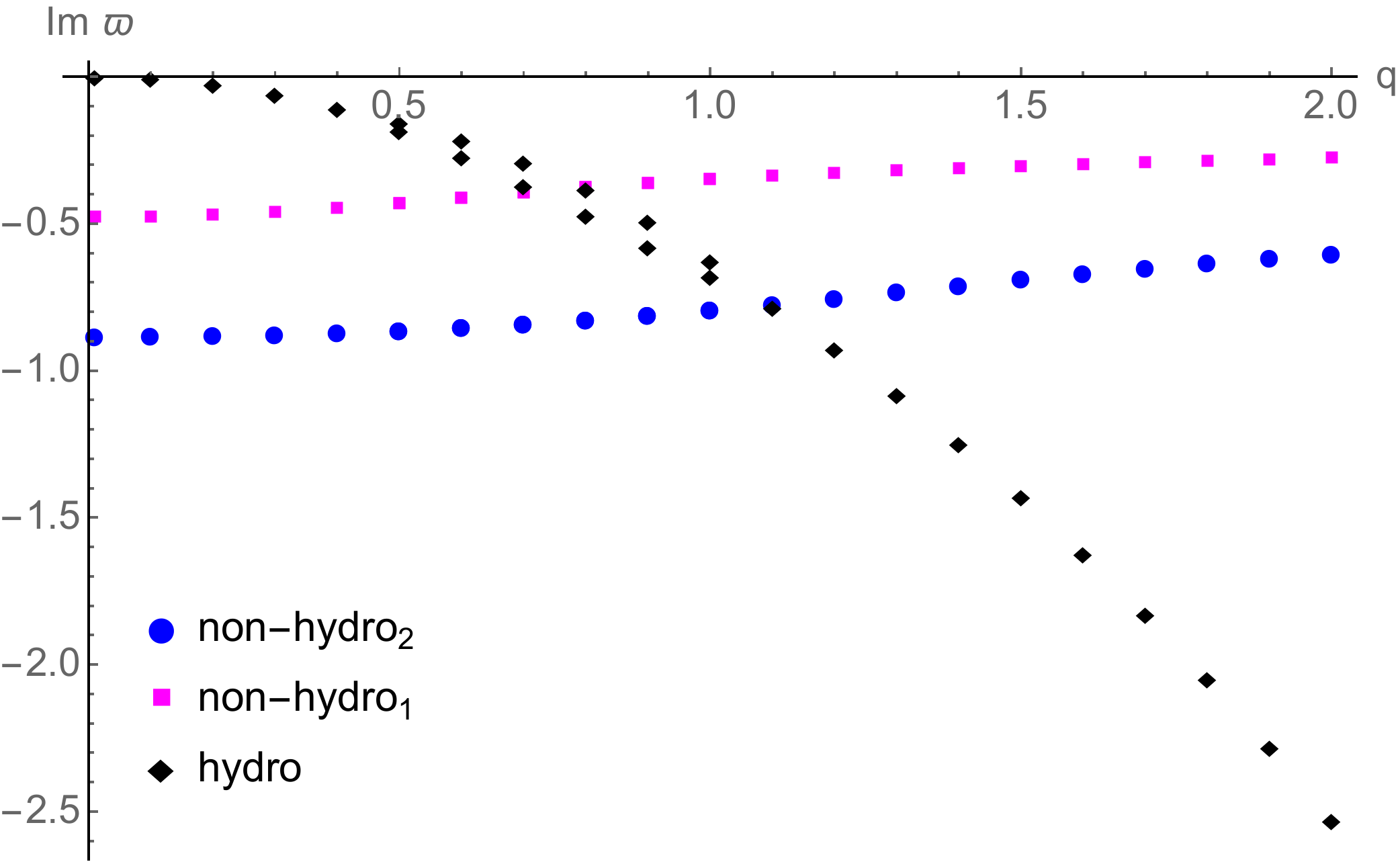}
	\caption{Quasinormal modes for the potential $V_{1\rm st}$ at $T\simeq1.00004T_m$.
		Real part (upper panel) and imaginary part (lower panel).
		}
	\label{V1st(QNM-2)}
\end{figure}

The second observation is the bubble formation in the spinodal 
region in the case of the first order phase transition \cite{Chomaz:2003dz}.
This happens when $c_s^2<0$ which means that hydrodynamic mode
is purely imaginary $\omega_H=\pm i |c_s| k - i \Gamma_sk^2$.
For small $k$, the mode with the plus sign is in the unstable region,
i.e., ${\rm Im} ~\omega_H>0$. For larger momenta the other term starts 
to dominate, so that there is $k_{\rm max}=|c_s|/\Gamma_s$ for which the hydro mode
becomes again stable. The scale of the bubble is the momentum for which
positive imaginary part of the hydro mode attains the maximal value.
Imaginary part of the unstable hydro mode is called the growth rate \cite{Chomaz:2003dz},
and is illustrated in the inset in fig.~\ref{V1st(QNM)}.
This phenomena is very similar to a Gregory-Laflamme instability \cite{Gregory:1993vy}.
When $\omega_H$ is purely imaginary, one can express it as
$\omega_H=\pm i O(k) - i E(k)$, with $O(-k)=-O(k)$ and $E(-k)=E(k)$.
Then there are two separated branches of the hydrodynamic modes,
as seen on fig. \ref{V1st(QNM)} and \ref{V1st(QNM-2)}.
In summary, in all cases considered, whenever there was an
indication of thermodynamic instability in equations of state,
the lowest QNM displayed a dynamical instability. Also,
nonhydrodynamic modes were always stable, even for large $k$. 
This supports the expectation spelled out in ref. \cite{Buchel:2005nt}.

The third observation is that near the minimal temperature
in the case of $1^{\rm st}$ order phase transition hydrodynamic
mode (cf. fig. \ref{V1st(QNM-2)}), and in the IHQCD case also the lowest nonhydrodynamic modes, become purely imaginary
for a range of momenta \cite{Janik:2016yt}.
The interpretation of this fact is that the corresponding wavelengths
cannot propagate at a linearized level, and  
there is a diffusion-like
mechanism for those modes. 
The onset of the appearance of a nonpropagating sound mode in the deeply overcooled phase has been observed earlier in a related model \cite{Gursoy:2013zxa}.
Our analysis shows that the range of momenta for which the nonpropagating mode appears is finite.
It is important to note that generically the ultra-locality \cite{Janik:2015waa}
of the nonhydrodynamic mode is still present in the critical region of the phase diagram.
The only exception observed is the IHQCD potential, where the modes exhibit a non
trivial behaviour \cite{Janik:2016yt}. 
Most of the interesting dynamics and effects observed are due to the
different behaviour of the hydrodynamic modes. This includes the instability and the 
bubble formation in the case of the $1^{\rm st}$ order phase transition.


\noindent {\bf  Conclusions.} In the present paper we performed an extensive study of the linearized
dynamics of excitations in strongly coupled field theories in the vicinity
of a nontrivial phase structure of various kinds.
We observed a number of novel features which were not present in the conformal case.
Firstly, for relatively small $k$, the propagating hydrodynamical sound modes become more damped than 
the lowest nonhydrodynamic degrees of freedom. This provides a more stringent restriction on the applicability
of hydrodynamics and indicates the necessity of incorporating these other degrees of freedom 
on appropriate length scales. This is in contrast to the conformal case where a similar phenomenon
only occurred in the shear channel and only at a higher value of $k$.
Secondly, we explicitly determined the instability in the spinoidal branch of a $1^{\rm st}$ order phase structure
and estimated the length scale for bubble formation.
Thirdly, close to the point $T=T_m$ on the $1^{\rm st}$ order equation of state, the sound mode
frequencies become purely imaginary for a range of momenta thus indicating that these modes effectively 
do not propagate at these length scales.
The richness of phenomena appearing in the linearized regime strongly suggests that it would be important to study the
corresponding real-time dynamics also at the nonlinear level.


\noindent{\bf Acknowledgments.} RJ and HS were supported by NCN grant 2012/06/A/ST2/00396, JJ by 
the NCN post-doctoral internship grant DEC-2013/08/S/ST2/00547.
We would like to thank D. Blaschke and P. Witaszczyk for interesting discussions.



\begin{thebibliography}{99}


\bibitem{Maldacena:1997re}
  J.~M.~Maldacena,
  Adv.\ Theor.\ Math.\ Phys.\  {\bf 2} (1998) 231, [hep-th/9711200].	
	
\bibitem{CasalderreySolana:2011us} 
  J.~Casalderrey-Solana {\it et al.}, 
  [arXiv:1101.0618 [hep-th]].


\bibitem{Kovtun:2005ev} 
  P.~K.~Kovtun and A.~O.~Starinets,
  Phys.\ Rev.\ D {\bf 72}, 086009 (2005),
  [hep-th/0506184].


\bibitem{Janik:2015waa} 
  R.~A.~Janik, G.~Plewa, H.~Soltanpanahi and M.~Spalinski,
  Phys.\ Rev.\ D {\bf 91}, no. 12, 126013 (2015),
  [arXiv:1503.07149 [hep-th]].


\bibitem{Buchel:2015saa} 
  A.~Buchel, M.~P.~Heller and R.~C.~Myers,
  Phys.\ Rev.\ Lett.\  {\bf 114}, no. 25, 251601 (2015),
  [arXiv:1503.07114 [hep-th]].


\bibitem{Janik:2016yt}
  R.~A.~Janik, J.~Jankowski and H.~Soltanpanahi 
  to appear.
  
\bibitem{Gursoy:2013zxa} 
U.~Gürsoy, S.~Lin and E.~Shuryak,
Phys.\ Rev.\ D {\bf 88}, no. 10, 105021 (2013)
[arXiv:1309.0789 [hep-th]].
  
   
\bibitem{Gubser:2008ny} 
  S.~S.~Gubser and A.~Nellore,
  Phys.\ Rev.\ D {\bf 78}, 086007 (2008),
  [arXiv:0804.0434  [hep-th]].


\bibitem{Gursoy:2008za} 
  U.~Gursoy, E.~Kiritsis, L.~Mazzanti and F.~Nitti,
  JHEP {\bf 0905}, 033 (2009),
  [arXiv:0812.0792  [hep-th]].

\bibitem{Benincasa:2005iv} 
  P.~Benincasa, A.~Buchel and A.~O.~Starinets,
  Nucl.\ Phys.\ B {\bf 733}, 160 (2006),
  [hep-th/0507026].

\bibitem{Heller:2013fn} 
  M.~P.~Heller, R.~A.~Janik and P.~Witaszczyk,
  Phys.\ Rev.\ Lett.\  {\bf 110}, no. 21, 211602 (2013),
  [arXiv:1302.0697  [hep-th]].

\bibitem{Gubser:2008yx} 
S.~S.~Gubser, A.~Nellore, S.~S.~Pufu and F.~D.~Rocha,
Phys.\ Rev.\ Lett.\  {\bf 101}, 131601 (2008),
[arXiv:0804.1950  [hep-th]].


\bibitem{Kovtun:2004de} 
  P.~Kovtun, D.~T.~Son and A.~O.~Starinets,
  Phys.\ Rev.\ Lett.\  {\bf 94}, 111601 (2005),
  [hep-th/0405231].

\bibitem{Janik:2006ft} 
  R.~A.~Janik,
  Phys.\ Rev.\ Lett.\  {\bf 98}, 022302 (2007),
  [hep-th/0610144].
	

\bibitem{Landsteiner:2012gn} 
  K.~Landsteiner,
  AIP Conf.\ Proc.\  {\bf 1458}, 174 (2011),
  [arXiv:1202.3550 [gr-qc]].


\bibitem{Chomaz:2003dz} 
  P.~Chomaz, M.~Colonna and J.~Randrup,
  Phys.\ Rept.\  {\bf 389}, 263 (2004),

\bibitem{Gregory:1993vy} 
  R.~Gregory and R.~Laflamme,
  Phys.\ Rev.\ Lett.\  {\bf 70}, 2837 (1993),
  [hep-th/9301052].




\bibitem{Buchel:2005nt} 
A.~Buchel,
Nucl.\ Phys.\ B {\bf 731}, 109 (2005),
[hep-th/0507275].




\end{thebibliography}
\end{document}